\documentclass[aps,pra,a4paper,10pt,superscriptaddress]{revtex4-1}
\usepackage{typearea}
\usepackage[american]{babel}
\usepackage[T1]{fontenc}
\usepackage[cp1252]{inputenc}
\usepackage{amsmath}
\usepackage[varg]{txfonts}
\usepackage{microtype}
\usepackage{graphicx}
\usepackage{units}
\usepackage{textcomp}
\usepackage{gensymb}

\newcommand{\TUKL}{Department of Physics and Research Center OPTIMAS, University of Kaiserslautern, 67663 Kaiserslautern, Germany}
\newcommand{\IPM}{Fraunhofer Institute for Physical Measurement Techniques IPM, 79110 Freiburg, Germany}
\newcommand{µ}{\micro}
\newcommand{°}{\degree}
\renewcommand{\pi}{\piup}
\newcommand{\eff}{\mathrm{eff}}
\newcommand{\epsr}{\epsilon_\mathrm{r}}
\newcommand{\im}{\mathrm{i}}
\renewcommand{\vec}{\mathbf}

\begin{document}

\title{Metamaterial near-field sensor for deep-subwavelength thickness measurements and sensitive refractometry in the terahertz frequency range}
\date{\today}
\author{Benjamin Reinhard}\email{breinhard@physik.uni-kl.de}\affiliation{\TUKL}
\author{Klemens M. Schmitt}\affiliation{\TUKL}
\author{Viktoria Wollrab}\affiliation{\TUKL}\affiliation{\IPM}
\author{Jens Neu}\affiliation{\TUKL}
\author{René Beigang}\affiliation{\TUKL}\affiliation{\IPM}
\author{Marco Rahm}\affiliation{\TUKL}\affiliation{\IPM}

\begin{abstract}
We present a metamaterial-based terahertz (THz) sensor for thickness measurements of subwavelength-thin materials and refractometry of liquids and liquid mixtures. The sensor operates in reflection geometry and exploits the frequency shift of a sharp Fano resonance minimum in the presence of dielectric materials. We obtained a minimum thickness resolution of \unit[12.5]{nm} (1/16000 times the wavelength of the THz radiation) and a refractive index sensitivity of \unit[0.43]{THz} per refractive index unit. We support the experimental results by an analytical model that describes the dependence of the resonance frequency on the sample material thickness and the refractive index.
\end{abstract}

\maketitle
\section{Introduction}

Terahertz (THz) radiation has attracted considerable attention during the last decades. This spectral range (frequencies between approx.\ \unit[0.1]{THz} and \unit[10]{THz}) is of special interest for sensing applications because many substances have a specific spectral response in this frequency interval \cite{fischer2005,markelz2008}.
The detection of small amounts or very thin layers of a sample material, however, remains a challenge in terahertz sensing technology. If the quantity of a sample material is too small, it may not cause a significant change of the detected signal in a simple transmission or reflection measurement. To increase the interaction of the terahertz field with the sample, several approaches have been proposed recently \cite{nagel2002,zhang2004,walther2005,rau2005,hasek2006,isaac2008,cheng2008,theuer2010,ohara2012}.
One particularly promising option to increase the sensitivity of terahertz measurements is the use of metamaterials, artificial materials whose electromagnetic response is determined mainly by the size and shape of subwavelength-sized inclusions. Due to the strong localization of the electromagnetic fields in the vicinity of resonant metallic elements, metamaterials exhibit a strong change of their optical response when a sample material is present. This property may be exploited to construct sensors which promise superior sensitivity. Examples of metamaterial-assisted sensing have been demonstrated in the microwave, terahertz, and near-infrared frequency ranges in the past years \cite{debus2007,driscoll2007,yoshida2007,o'hara2008,al-naib2008,lahiri2009,chiam2009,liu2010a,tao2010}.

Metamaterial-based sensors offer some important advantages over standard terahertz time-domain spectroscopy (THz-TDS).
With THz-TDS, the frequency-dependent complex refractive index of a sample under test can be determined from a reflection or transmission measurement. However, when calculating optical constants from terahertz measurements, one has to take great care concerning the influence of signal noise on the calculated values \cite{withayachumnankul2008}. Especially the noise of measured amplitudes has a great influence on the derived data. Furthermore, in transmission geometry, the thickness of the sample has to be known exactly. In reflection geometry, an additional uncertainty is the phase of the reference spectrum.
These limitations can be overcome by metamaterial-based sensors which allow to derive the refractive index and the thickness of a sample material from frequency measurements rather than from amplitude measurements. As a main advantage over amplitude measurements, frequency measurements are less prone to noise and do not require the knowledge of a reference spectrum, provided that the spectrum of the terahertz source is smooth in the spectral range of interest.

In this letter, we present a metamaterial-based sensor whose reflection shows a strong frequency shift of a Fano-type resonance minimum in the presence of a dielectric sample. The magnitude of this shift depends on both the refractive index and the thickness of the sample. The metamaterial is designed to operate at frequencies between approx.\ \unit[1]{THz} and \unit[1.7]{THz}. This range is easily accessible with standard terahertz sources and detectors.

\section{Metamaterial design}

\begin{figure}%
\centering%
\includegraphics{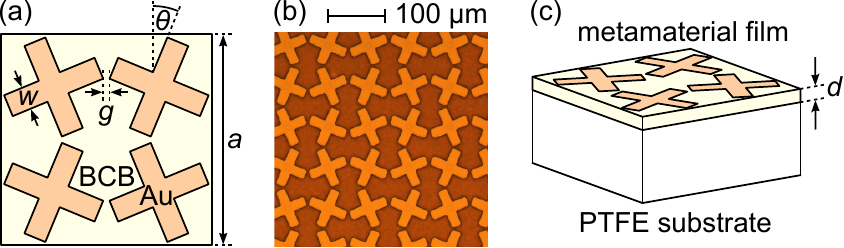}%
\caption{(a) Geometry parameters of the metamaterial unit cell. $a=\unit[140]{µm}$, $w=\unit[15]{µm}$, $g=\unit[4]{µm}$, $\theta=22.5°$. The metamaterial consists of gold (Au) crosses on top of a benzocyclobutene (BCB) film. (b) Microscope image of a fabricated metamaterial. (c) The $d=\unit[10]{µm}$ thick metamaterial film is glued on a \unit[5]{mm} thick PTFE substrate if mechanical rigidity is required.}%
\label{fig:design}%
\end{figure}%

\begin{figure}%
\centering%
\includegraphics{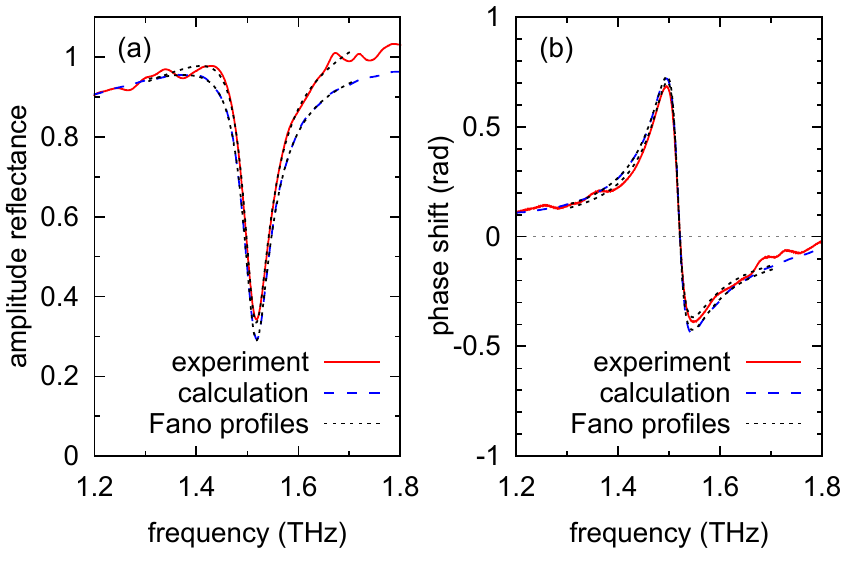}%
\caption{(a) Terahertz amplitude and (b) phase of the reflection spectrum of the sensor without any sample material. Both the experimental and the numerical data follow a Fano profile.}%
\label{fig:spectrum}%
\end{figure}

The unit cell of the metamaterial consists of four metallic crosses on top of a \unit[10]{µm} thick dielectric matrix with $\epsr=2.67$. The unit cell is square with an edge length of \unit[140]{µm}.
Each of the crosses is tilted by an angle of 22.5°. The geometry parameters are shown in detail in Fig.~\ref{fig:design}(a).
Despite the relatively large unit cell, diffraction by the metamaterial occurs only for frequencies greater than approx.\ \unit[2.1]{THz} because the first order of diffraction is zero due to the symmetry of the structure.
Figures~\ref{fig:spectrum}(a) and \ref{fig:spectrum}(b) show the experimentally measured terahertz reflection along with corresponding numerical calculations. The spectrum shows a pronounced, narrow reflection minimum.
The position of the reflection minimum does not change when the angle of incidence of the incoming terahertz radiation is changed within $\pm 30°$ as long as transverse magnetic (TM) polarization is maintained.
The advantage of operation of the sensor in reflection geometry is that the sample only has to be accessible from one side, which is especially important when measuring liquids. Furthermore, in contrast to transmission measurements, the absorption of the sample does not reduce the signal amplitude when measuring reflection. A narrow resonance feature is necessary to accurately determine the magnitude of a frequency shift and thus increases the resolution of the sensor material. In our design, the spectral width $\Delta\omega$ (full half-width) of the reflection minimum with respect to the resonance frequency $\omega_0$ is $\Delta\omega/\omega_0\approx 1/26$.

In the region around the resonance frequency, the reflection spectrum displays a Fano profile \cite{luk'yanchuk2010}. The function
\begin{equation}%
\vec{\tilde E}(\omega) = \frac{\tilde q\varGamma/2 + \omega-\omega_0}{\omega-\omega_0+\im\varGamma/2}~\vec{\tilde E_0}(\omega)
\end{equation}%
fits the frequency-dependent complex field amplitude of the experimental and numerical data (see Figs.~\ref{fig:spectrum}(a) and \ref{fig:spectrum}(b)), where $\vec{\tilde E_0}(\omega)$ is a baseline with linear dependence of the magnitude and the phase, $\varGamma$ is the width and $\omega_0$ the angular frequency of the resonance. The complex Fano parameter $\tilde q$ describes the asymmetry of the resonance profile and the amount of dissipation in the system \cite{baernthaler2010}, which is directly related to the depth of the resonance minimum.

\section{Analytical model}

\begin{figure}%
\centering%
\includegraphics{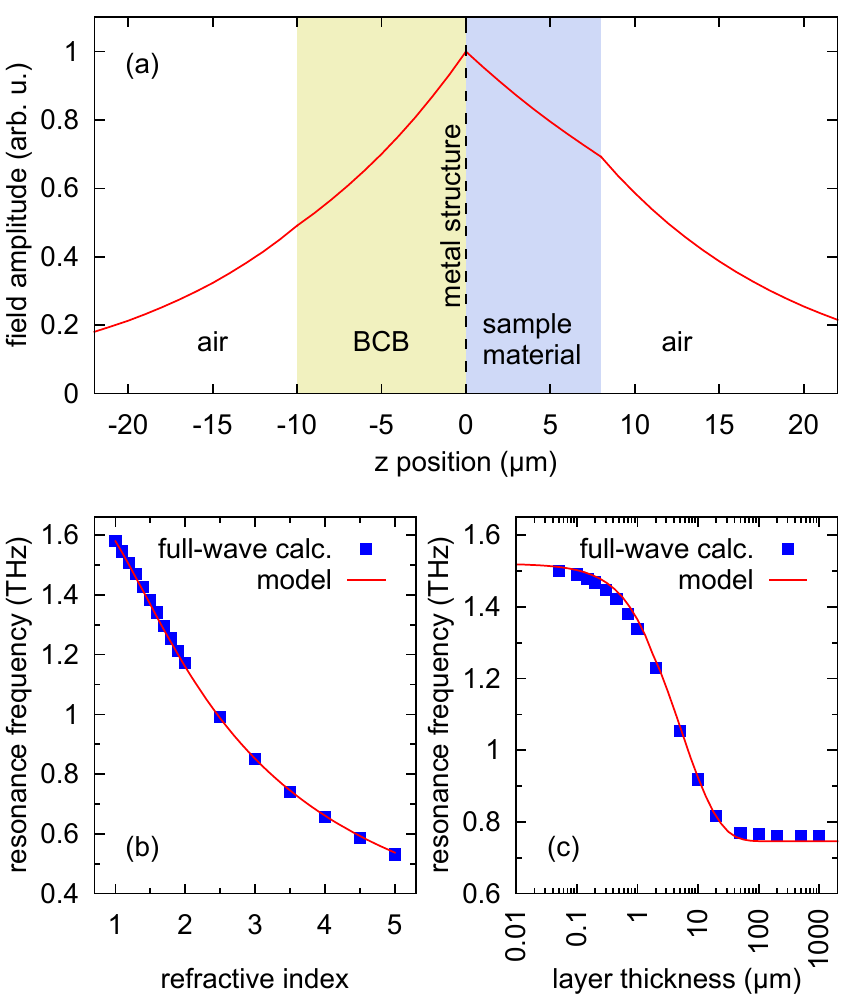}%
\caption{(a) The metamaterial sensor and the sample material form a layered system with the $z$ direction as the normal on the layers. The field amplitude decreases exponentially from the metal structure at $z=0$. The wave vector $k_z(z)$ depends on the permittivities of the layers.
(b) Resonance frequencies of the sensor in the presence of a sample material with variable refractive index and infinite thickness and (c) variable thickness and a constant refractive index of $n=3.42$. The agreement between the full-wave numerical calculations and the analytical model is very good. Note the logarithmic scaling of the horizontal axis in (c).}%
\label{fig:model}%
\end{figure}

The shift of the resonance frequency in the presence of a sample material can be explained using an intuitive model which describes the sensor and the sample material as a layered system of homogeneous media (Fig.~\ref{fig:model}(a)). The normal on the layer surfaces is defined as the $z$ direction. The metamaterial sensor itself can be understood as an LC circuit. Its resonance frequency $\omega_0=(L_\eff \times C_\eff)^{-\nicefrac{1}{2}}$ is determined by an effective inductance $L_\eff$ and an effective capacitance $C_\eff$. Because we only consider nonmagnetic sample materials with a relative permeability $\mu_\mathrm{r}=1$, the effective inductance of the system is expected to be independent of the sample material. The effective capacitance, however, is influenced by the relative permittivities of the sensor and the sample material and can be described as
\begin{equation}%
C_\eff = \frac{Q_\eff}{U_\eff} = \frac{\int_V\epsilon_0\epsilon_\mathrm{r}\vec E\cdot \mathrm{d}\vec A}{\int_{P_1}^{P_2}\vec E \cdot \vec{\hat x}~\mathrm{d}x} \propto \int \epsr(z)\vec E(z)\cdot \vec{\hat x}~dz~.
\label{eq:Ceff}%
\end{equation}%
Note that $\epsr(z)$ reflects the spatial dependence of the relative permittivity in the sensor and the sample material along the normal on the layers.
For the $z$ dependence of the electric field $\vec E(z)$ we assume that the electric field amplitude decreases exponentially with increasing distance from the metallizations as shown in Fig.~\ref{fig:model}(a). The wave number $k_z$ which describes this exponential decrease is determined by diffraction at the periodic structure, where we consider only the first non-zero Fourier component of the structure with a transverse wave number $k_\parallel$:
\begin{equation}
k_z(z) = \sqrt{\epsr(z)\left(\frac{\omega}{c_0}\right)^2-k_\parallel^2}
\end{equation}
For the investigated design, the wave number is $k_\parallel = 2\pi/(\unit[70]{µm})$. Note that $k_z$ depends on the permittivity of the dielectric layers and thus possesses a $z$ dependence. Reflections at the boundaries between the layers are neglected.

While we cannot reliably quantify the absolute value of the effective capacitance in Eq.~\eqref{eq:Ceff} without knowing the exact field distribution of the resonant structure, we can calculate a relative effective capacitance for each layered system of materials surrounding the plane of the metallizations. This enables us to calculate the frequency shift in the presence of a sample material. The resonance frequency $\omega_\mathrm{s}$ with a sample material on top of the sensor is then
\begin{equation}
\omega_\mathrm{s} = \omega_0 \sqrt{\frac{C_\mathrm{eff}(\omega_0)}{C_\mathrm{eff}(\omega_\mathrm{s})}}~,
\end{equation}
where $\omega_0$ is the resonance frequency without the sample material. Because $C_\mathrm{eff}$ depends on the frequency, we employ an iterative method to determine $\omega_\mathrm{s}$.

In order to test the validity of this simple model, we performed two series of full-wave numerical calculations using \emph{CST Microwave Studio\textsuperscript\textregistered}. In the first series, we calculated the reflection spectrum and the resonance frequency of the sensor in the presence of thick layers of dielectric sample materials with a refractive index between 1 and 5 ($\epsr$ between 1 and 25). The thickness of the sample material was set to \unit[300]{µm}. At this distance, the evanescent fields at resonance frequency have decayed by a factor of more than $10^9$, which means that the sample materials appeared infinitely thick to the sensor. In the second set of calculations, we investigated the resonance frequencies of the sensor with dielectric layers with constant refractive index $n=3.42$ and thicknesses between \unit[50]{nm} and \unit[1]{mm}. Figures~\ref{fig:model}(b) and \ref{fig:model}(c) show the dependence of the resonance frequencies on the refractive index and the thickness of the sample material. In both graphs we compared the resonance frequencies obtained from full-wave calculations with the predicted values of the model. Because the model provides no information about the absolute value of the resonance frequency, the resonance frequency without sample material was set to match the value derived from the full-wave calculations. Within the investigated range of refractive indices and layer thicknesses of the sample materials, the resonance frequencies of the model and the numerical calculations agreed very well. The remaining small deviations can be explained by the fact that we neglected more strongly decaying near field components with larger transverse wave numbers.

\section{Fabrication and measurement}

We fabricated the metamaterial sensors by UV lithography. We used a \unit[10]{µm} thick layer of benzocyclobutene (BCB) as a dielectric background material. On top of this layer, we evaporated a \unit[200]{nm} thick gold structure. We then removed the silicon substrate which we used during the fabrication process. Fig.~\ref{fig:design}(b) shows a microscope image of a fabricated sensor. Details on the fabrication process can be found in an earlier publication\cite{paul2008}.

A requirement on the metamaterial design was robustness against deviations of the structure parameters which may occur during the fabrication process. A slight variation is expected because of the sensitivity of the lithography process with respect to the exposure and development times. We estimated the standard deviation of the resonance frequencies of individual sensors to be smaller than \unit[5]{GHz}.
This means that imperfections and deviations from the nominal design which arise during the fabrication process pose virtually no restrictions on the accuracy of the frequency measurement.

We optically characterized and tested the performance of the sensors by measuring the reflection spectra using THz time-domain spectroscopy. We employed photoconductive switches as emitter and detector. By using a \unit[10]{mm} thick high-resistivity silicon wafer as a beam splitter, we measured the reflection spectra of the sensors under normal incidence with a spectral resolution of approx.\ \unit[5]{GHz}. We took all measurements in a dry air atmosphere (relative humidity $<\unit[4]{\%}$) at room temperature (\unit[21]{°C}).

\section{Experimental results}

\begin{figure}%
\centering%
\includegraphics{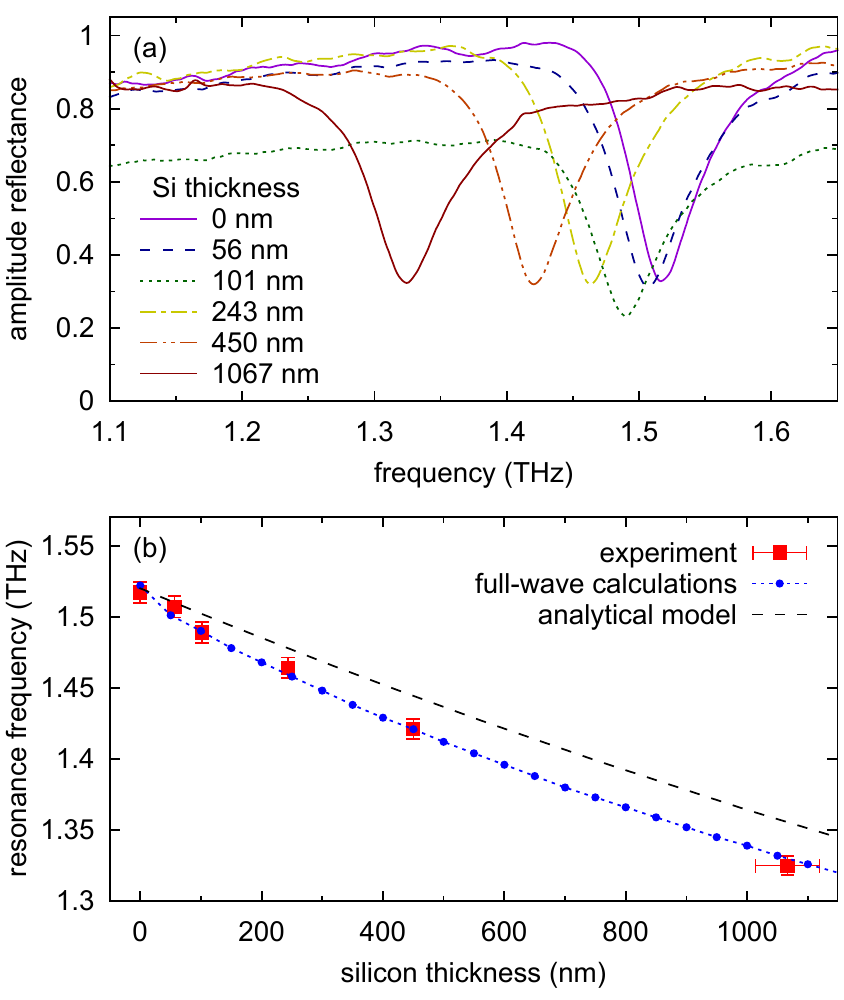}%
\caption{(a) Experimentally measured reflection spectra of the sensor with different thicknesses of silicon layers. A thicker layer lowers the resonance frequency of the sensor. (b) Resonance frequencies of the sensor in dependence on the silicon thickness. The slope of the curve has a maximum value of \unit[0.4]{THz/µm}.}%
\label{fig:thin_film_exp}%
\end{figure}

In order to experimentally prove the capability of the sensor of measuring the thickness of thin sample materials, we evaporated silicon layers ($n \approx 3.4$) with thicknesses between approx.\ \unit[50]{nm} and \unit[1]{µm} on top of the metamaterial sensor. Prior to the THz measurements, we measured the thickness of the silicon layers using a surface profilometer. To assist in the fabrication process, we covered the silicon with an additional sub-\unit[10]{µm} thick layer of BCB. This created an additional frequency shift but did not influence the functionality of the sensor.
Figures~\ref{fig:thin_film_exp}(a) and \ref{fig:thin_film_exp}(b) show the variation of the resonance frequency of the sensor for different thicknesses of silicon. As expected, a thicker silicon layer lowers the resonance frequency of the sensor. We observed frequency shifts in a range from \unit[1.52]{THz} without silicon to \unit[1.32]{THz} for a silicon thickness of \unit[1070]{nm}.
The measured frequency shifts agree very well with both the numerical calulations and the predictions of the analytical model.
From Fig.~\ref{fig:thin_film_exp}(b) we read that the sensitivity of the sensor has a maximum value of approx.\ \unit[0.4]{THz/µm} for very small layer thicknesses. Assuming a frequency resolution of \unit[5]{GHz}, this yields a thickness resolution of \unit[12.5]{nm}, which corresponds to approx.\ 1/16000 of the terahertz wavelength ($\lambda \approx \unit[200]{µm}$).

\begin{figure}%
\centering%
\includegraphics{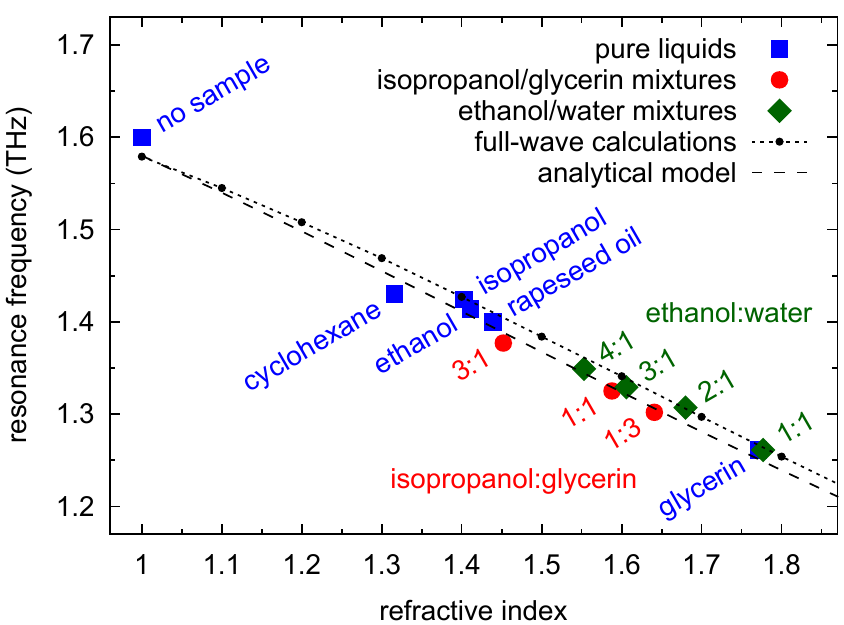}%
\caption{Experimentally measured resonance frequencies of the metamaterial sensor in the presence of several liquids and liquid mixtures. The data of butanol, paraffin, and NEP lie very close to the data of ethanol and isopropanol and have been omitted from the graph for the sake of clarity. The data points from the measurements of the mixtures are labeled with the volume ratio of the constituents.}%
\label{fig:ref_index_measurements}%
\end{figure}

In a second series of experiments, we investigated the influence of different liquids and liquid mixtures on the resonance frequency of the metamaterial sensor. We brought the liquids in direct contact with the metallic structure of the sensor.
To provide the sensor with higher mechanical stability, we glued it on a \unit[5]{mm} thick PTFE substrate (Fig.~\ref{fig:design}(c)). We deliberately chose PTFE as a substrate because of its low refractive index ($n\approx 1.4$) and low absorption at terahertz frequencies. A substrate with a high index of refraction (such as silicon with $n\approx 3.4$) would have considerably reduced the sensitivity of the sensor \cite{o'hara2008,tao2010}.
The liquids under investigation were isopropanol, glycerin, paraffin, ethanol, butanol, rapeseed oil, cyclohexane, and N-ethyl-2-pyrrolidone (NEP).
In addition to the pure liquids, we measured the reflection spectrum of the sensor for different mixtures of isopropanol and glycerin as well as ethanol and water. 
To calibrate the metamaterial sensor, we determined the refractive indices of the liquids and liquid mixtures in an alternative experimental configuration. For this purpose, we measured the reflection spectra of the liquids through a silicon window. For each liquid or liquid mixture, we obtained the reflection spectra at the air/silicon interface and the silicon/liquid interface. From this information, we calculated the refractive indices of the liquids \cite{thrane1995}. This technique can be applied for any sample material with weak dispersion which does not imply a significant increase of the terahertz pulse length. Based on these measurements, we established the relation between the resonance frequency of the sensor and the refractive index of the liquids (Fig.~\ref{fig:ref_index_measurements}).
In a refractive index range between 1 and 1.8, the dependence between resonance frequency and refractive index is almost linear with a slope of approx.\ $\unit[-0.43]{THz}$ per refractive index unit.
Assuming an error of \unit[5]{GHz} for the resonance frequency, we could reliably detect refractive index differences as small as 0.01 with the metamaterial sensor. It should be mentioned that such a high resolution can only be achieved for weakly absorbing sample materials. In general, the resolution of the sensor decreases with increasing loss in the sample material. For all investigated sample materials, however, the refractive index resolution was better than 0.09.

\section{conclusion}

In conclusion, we presented a metamaterial-based terahertz (THz) sensor for the measurement of thin sample materials with subwavelength thickness and for refractive index sensitive measurements of liquids and liquid mixtures. As a measurement signal we exploited the frequency shift of a sharp Fano resonance in the reflection spectrum of the metamaterial in the presence of a dielectric. The sensor has been devised such that the sample material under investigation faces the metallic structure of the metamaterial while the THz beam is incident from the opposite side. As a major advantage of operation in reflection geometry, the THz beam need not be transmitted through the sample material, thus ensuring low absorption and an increase of the signal-to-noise ratio. For silicon as a sample material we observed a frequency shift of \unit[0.4]{THz/µm}, which corresponds to a thickness resolution of \unit[12.5]{nm} (1/16000 times the wavelength of the THz radiation). Furthermore, we determined the refractive index of various liquids and liquid mixtures with different mixing ratios. With the sensor, we obtained a refractive index resolution of up to 0.01 by measuring the resonance shift of the detector with a slope of $\unit[-0.43]{THz}$ per refractive index unit. We explained the physical behavior of the detector by an analytical model which predicted the correct dependence of the resonance frequency shift on the sample material thickness and the refractive index.

\end{document}